\documentclass[a4paper,12pt]{article}
\usepackage{amsfonts}
\usepackage{latexsym}
\usepackage{amssymb}
\date{}

\begin{document}

\title{Mathisson-Papapetrou Equations\\as Conditions for the Compatibility
of\\General Relativity and Continuum Physics}
\author{W. Muschik\quad and\quad 
H.-H. v. Borzeszkowski\footnote{Corresponding author: 
borzeszk@itp.physik.tu-berlin.de}
\\
Institut f\"ur Theoretische Physik\\
Technische Universit\"at Berlin\\
Hardenbergstr. 36\\D-10623 BERLIN,  Germany}
\maketitle

            \newcommand{\be}{\begin{equation}}
            \newcommand{\beg}[1]{\begin{equation}\label{#1}}
            \newcommand{\ee}{\end{equation}\normalsize}
            \newcommand{\bee}[1]{\begin{equation}\label{#1}}
            \newcommand{\bey}{\begin{eqnarray}}
            \newcommand{\byy}[1]{\begin{eqnarray}\label{#1}}
            \newcommand{\eey}{\end{eqnarray}\normalsize}
            \newcommand{\beo}{\begin{eqnarray}\normalsize}
            \newcommand{\R}[1]{(\ref{#1})}
            \newcommand{\C}[1]{\cite{#1}}

            \newcommand{\mvec}[1]{\mbox{\boldmath{$#1$}}}
            \newcommand{\x}{(\!\mvec{x}, t)}
            \newcommand{\m}{\mvec{m}}
            \newcommand{\F}{{\cal F}}
            \newcommand{\n}{\mvec{n}}
            \newcommand{\argm}{(\m ,\mvec{x}, t)}
            \newcommand{\argn}{(\n ,\mvec{x}, t)}
            \newcommand{\T}[1]{\widetilde{#1}}
            \newcommand{\U}[1]{\underline{#1}}
            \newcommand{\X}{\!\mvec{X} (\cdot)}
            \newcommand{\cd}{(\cdot)}
            \newcommand{\Q}{\mbox{\bf Q}}
            \newcommand{\p}{\partial_t}
            \newcommand{\z}{\!\mvec{z}}
            \newcommand{\bu}{\!\mvec{u}}
            \newcommand{\rr}{\!\mvec{r}}
            \newcommand{\w}{\!\mvec{w}}
            \newcommand{\g}{\!\mvec{g}}
            \newcommand{\D}{I\!\!D}
            \newcommand{\se}[1]{_{\mvec{;}#1}}
            \newcommand{\sek}[1]{_{\mvec{;}#1]}}            
            \newcommand{\seb}[1]{_{\mvec{;}#1)}}            
            \newcommand{\ko}[1]{_{\mvec{,}#1}}
            \newcommand{\ab}[1]{_{\mvec{|}#1}}
            \newcommand{\abb}[1]{_{\mvec{||}#1}}
            \newcommand{\td}{{^{\bullet}}}
            \newcommand{\eq}{{_{eq}}}
            \newcommand{\eqo}{{^{eq}}}
            \newcommand{\f}{\varphi}
            \newcommand{\dm}{\diamond}
            \newcommand{\seq}{\stackrel{_\bullet}{=}}
            \newcommand{\st}[2]{\stackrel{_#1}{#2}}
            \newcommand{\om}{\Omega}
            \newcommand{\emp}{\emptyset}
            \newcommand{\bt}{\bowtie}
            \newcommand{\btu}{\boxdot}
\newcommand{\Section}[1]{\section{\mbox{}\hspace{-.6cm}.\hspace{.4cm}#1}}
\newcommand{\Subsection}[1]{\subsection{\mbox{}\hspace{-.6cm}.\hspace{.4cm}
\em #1}}

%\date{\today}
%\tableofcontents
%eigene makros
\newcommand{\const}{\textit{const.}}
\newcommand{\vect}[1]{\underline{\ensuremath{#1}}}  %Vektoren
\newcommand{\abl}[2]{\ensuremath{\frac{\partial #1}{\partial #2}}}

\abstract\noindent
In continuum physics is presupposed that general-relativistic balance equations
are valid which are created from the Lorentz-covariant ones by application
of the equi\-valence principle. Consequently, the question arises, how to make
these general-covariant balances compatible with Einstein's field equations.
The compatibility conditions are derived by performing a modified
Belinfante-Rosenfeld symmetrization for the non-symmetric and not
divergence-free general-relativistic energy-mo\-men\-tum tensor. The procedure
results in the Mathisson-Papapetrou equations.

\section*{ }

In General Relativity Theory (GRT), as a consequence of Einstein's equations
\bee{b1}
R^{ab} - \frac{1}{2}g^{ab}R\ =\ \kappa\Theta^{ab}\quad\Longrightarrow\quad
\Theta^{ab}\ =\ \Theta^{ba},\quad
\Theta^{ab}_{\ \ ;b}\ =\ 0,
\ee
the energy-momentum tensor $\Theta^{ab}$ has to be symmetric and
divergence-free\footnote{We use the definitions of the curvature and Ricci
tensors given in \C{1}; the comma denotes partial and the semicolon
covariant derivatives.}. Starting out with an action principle for deriving
\R{b1}, $\Theta$ is the metric energy-momentum tensor\footnote{This tensor can
also be derived by exploiting the properties of the diffeomorphism group
\C{2}.} defined by
\bee{b2} 
\Theta^{ab}\ :=\ \frac{2}{\sqrt{-g}}
\frac{\delta {\cal L}_{mat}}{\delta g_{ab}},\qquad\mbox{with}\quad
{\cal L}_{mat}\ =\ {\cal L}_{mat}(g_{ab,c},\Phi^A,\Phi^A_{,a}),
\ee
the Lagrange density of that matter-field $\Phi^A$ which is the source of the
gravitational field described by the metric $g_{ab}$, determined by the
solution of \R{b1}. 
To be in accordance with the Einstein principle of
equivalence, the general-covariant tensor $\Theta^{ab}$ should recover the 
corresponding Lorentz-covariant energy-mo\-men\-tum tensor of the matter in a
local-geodesic coordinate system. Only symmetric
and divergence-free Lorentz-covariant energy-momentum tensors can be
transfered to a general-covariant tensor which can be used in the field
equations \R{b1} as a source of matter. 
\vspace{.3cm}\newline%%%%%%%%%%%%%%%%%%%%%%%%%%%%%%%%%%
One starts with a Lorentz-covariant canonical energy-momentum tensor known
from Special Relativity Theory (SRT) stemming from 
${\cal L}_{mat}(\eta_{ab},\Phi^A,\Phi^A_{,a})$ 
\bee{b3}
c\hspace{-.2cm}T^a_{\ b}\ :=\ 
\frac{\partial {\cal L}_{mat}}{\partial \Phi^A_{,a}}
\Phi^A_{,b} - \delta^a_b{\cal L}_{mat}
\ee
which in general is non-symmetric\footnote{This tensor is constructed by
applying the Noether theorem. In \C{3}, it is argued that the Noether
procedure can also be performed resulting in a symmetric tensor. In this case,
a symmetrization procedure is not needed.} and divergence-free
\bee{b4}
c\hspace{-.2cm}T^{ab}\ \neq\ c\hspace{-.2cm}T^{ba},\qquad
c\hspace{-.2cm}T^{ab}_{\ \ ,a}\ =\ 0.
\ee
This tensor has to be symmetrized. Without symmetrization procedure,
$c\hspace{-.2cm}T^{ab}$ could not be used as matter-source term on the
right-hand side of \R{b1}. 
\vspace{.3cm}\newline%%%%%%%%%%%%%%%%%%%%%%%%%%%%%%%%%%
If matter-field equations can be derived by an action principle, balance
equations for the spin are implied by the Bianchi identities\footnote{see,
e.g., references \C{4,5}}. In this sense, energy-momentum and spin balance
are dependent on each other: they stem from the same origin. Here, our point
of view is more pragmatic and less axiomatic: 
it concerns the fact that the phenomenological realm of application of
GRT is wider than that one sketched above\footnote{If there is no Lagrange
density --perhaps unknown or not existing-- in continuum physics 
energy-momentum balance and spin
balance have to be formulated separately and independently of each other.}.
Often, one has neither a matter Lagrangian (or another specified matter model)
nor does the energy-momentum tensor satisfy the conditions \R{b1}$_{2,3}$.
Then, in a special-relativistic version, one has phenomenological balance
equations of the following type for energy-momentum and spin
\bee{b5}
T^{bc}_{\ \ ,b}\ =\ k^c,\ \qquad\mbox{with}\quad T^{bc}\ \neq\ T^{cb}, 
\footnotemark[6]
\ee
and
\bee{b5o}
S^{cba}_{\ \ \  ,c}\ =\ m^{ba}\qquad\mbox{with}\quad S^{cba}\ =\ -S^{cab}\quad
\mbox{and}\quad m^{ba}\ =\ -m^{ab}.
\ee
\footnotetext[6]{If there is a Lagrange density for the considered matter,
then one has $T^{ab} = c\hspace{-.2cm}T^{ab}$.} 
For non-isolated systems, $k^c\neq 0$ denotes an external force
density, $m^{[ab]}$ is an external momentum density, and $S^{cba}$ the
current of spin density\footnote[7]{often shortly denoted as spin tensor}.
In particular, one finds such a situation in special-relativistic continuum
thermodynamics, where the balances \R{b5} must be supplemented by the balance
equations of particle number and entropy density\footnote[8]{For this continuum
theory of irreversible processes, see the contributions in \C{4} and \C{5a}.}.
To consider these balance equations within GRT one has to rewrite them in a
general-covariant form:
\bee{b5a}
T^{bc}_{\ \ ;b}\ =\ k^c,\ \qquad S^{cba}_{\ \ \ \ ;c}\ =\ m^{ba},
\ee
Now, the question arises: How can these balance equations be incorporated
beside the gravitational equations \R{b1} into the general-covariant framework
of GRT, without getting in contradiction\footnote[9]{as e.g. assumed in \C{7}}?
In the present brief note, we ask for these conditions by which a
general-relativistic construction of a symmetric energy-momentum tensor from
the non-symmetric $T^{bc}$ is possible, so that in a local-geodesic coordinate
system the relation between these tensors reduces to their special-relativistic
relation given by the Belinfante-Rosenfeld symmetrization \C{8}.
\vspace{.3cm}\newline%%%%%%%%%%%%%%%%%%%%%%%%%%%%%%%%%%%%%%%
We start out for remembering with a sketch of the usual Belinfante-Rosenfeld
symmetrization applied to the canonical energy-momentum tensor \R{b4} \C{8}.  
Defining a hyper-potential $\Sigma^{abc}$
\bee{b7}
\Sigma^{abc}\ :=\ 
S^{abc} + S^{bca} + S^{cba},\qquad S^{abc}\ =\ -S^{acb}\ \footnotemark[10]
\qquad\Sigma^{abc}\ =\ -\Sigma^{bac}.
\ee
\footnotetext[10]{Here $S^{abc}$ is given by the Lagrange density
${\cal L}_{mat}(\eta_{ab},\Phi^A,\Phi^A_{,a})$.}
Belinfante and Rosenfeld define the tensor
\bee{b9}
B^{bc}\ :=\ c\hspace{-.2cm}T^{bc} - \frac{1}{2}\Sigma^{abc}_{\ \ \ ,a}.
\ee
Since in SRT by use of the Lagrange density \R{b2}$_2$, $S^{abc}$ is fixed
and
\bee{b9a}
\Sigma^{a[bc]}_{\ \ \ ,a}\ \equiv\
S^{abc}_{\ \ ,a}\ =\ 2\ c\hspace{-.2cm}T^{[bc]} 
\quad\Longrightarrow\quad B^{[bc]}\ =\ 0 
\ee
holds true, $B^{bc}$ is symmetric. Furthermore, because of the vanishing
divergence of the canonical energy-momentum tensor and because of \R{b7}$_3$
and commuting partial derivatives, one obtains
\bee{b13}
\Sigma^{abc}_{\ \ \ ,a,b}\ =\ 0\quad\Longrightarrow\quad 
B^{bc}_{\ \ ,b}\ =\ 0
\ee
that the divergence of $B^{bc}$ vanishes. Consequently, one obtains the desired
special-relativistic relations
\bee{b14} 
B^{[bc]}\ =\ 0,\qquad B^{bc}_{\ \ ,b}\ =\ 0.
\vspace{.3cm}\ee%%%%%%%%%%%%%%%%%%%%%%%%%%%%%%%%%%%%%%%
We now investigate, if such a symmetrization procedure can also operate in GRT.
In contrast to the usual procedure, we do not take the general-covariantly 
rewritten symmetrized tensor $B^{bc}$ in \R{b9}, satisfying \R{b14}, as
source of Einstein's equations, but we set out with the
general-covariantly rewritten full tensor \R{b5}$_1$, that means, with $T^{bc}$
instead of $c\hspace{-.2cm}T^{bc}$.
Analogously to \R{b7}, we make the following ansatz
\byy{b16}
{^\dagger}\Sigma^{abc} &:=& 
{^\dagger}S^{abc} +{^\dagger} S^{bca} +{^\dagger} S^{cba},
\\ \label{b16a}
{^\dagger}S^{abc} &=& -{^\dagger}S^{acb},\qquad
{^\dagger}\Sigma^{abc}\ =\ -{^\dagger}\Sigma^{bac}.
\eey
Except for the anti-symmetry in the two last indices, ${^\dagger}\!S^{abc}$ is
not specified so far. Then, motivated by the Belinfante-Rosenfeld procedure,
we define
\bee{b15}
^\dagger\! B^{bc}\ :=\ T^{bc} - \frac{1}{2}{^\dagger} \Sigma^{abc}_{\ \ \ ;a},
\ee
Because this tensor does not automatically satisfy \R{b14},
we now have to demand that $^\dagger\!B^{bc}$ has to be symmetric and
divergence-free:
\byy{b17}
^\dagger B^{[bc]}\ \st{\td}{=}\ 0&\quad\Longrightarrow\quad&
^\dagger \Sigma^{a[bc]}_{\ \ \ ;a}\ \equiv\ 
^\dagger S^{abc}_{\ \ \ ;a}\ =\ 2T^{[bc]},
\\ \label{b17a}
^\dagger B^{bc}_{\ \ \ ;b}\ \st{\td}{=}\ 0&\quad\Longrightarrow\quad&
T^{bc}_{\ \ \ ;b} - \frac{1}{2}{^\dagger}\!\Sigma^{abc}_{\ \ \ ;a;b}\ =\ 0.
\vspace{.3cm}\eey%%%%%%%%%%%%%%%%%%%%%%%%%%%%%%%%%%%
For calculating ${^\dagger}\Sigma^{abc}_{\ \ \ ;a;b}$, we
start out with the relation for the second covariant derivative taking the
symmetry properties of $\Sigma^{abc}$ and those of the curvature tensor
$R^a_{\ bc}$ into account
\bey\nonumber
2{^{\dagger}}\!\Sigma^{abc}_{\ \ \ ;\,a;\,b} &=&
^{\dagger}\!\Sigma^{abc}_{\ \ \ ;\,a;\,b} - 
^{\dagger}\! \Sigma^{abc}_{\ \ \ ;\,b;\,a}\ =\
\\ \nonumber
&=&
R^a_{mab}{^{\dagger}}\! \Sigma^{mbc} + R^b_{mab}{^{\dagger}}\! \Sigma^{amb} + 
R^c_{mab}{^{\dagger}}\! \Sigma^{abm}\ =\
\\ \label{b18}
&=&
R_{mb}{^{\dagger}}\! \Sigma^{mbc} - R_{ma}{^{\dagger}}\! \Sigma^{amb} + 
R^c_{mab}{^{\dagger}}\! \Sigma^{abm}\ =\ R^c_{mab}{^{\dagger}}\! \Sigma^{abm}.
\eey
This results in
\bey\nonumber
2^{\dagger}\!\Sigma^{abc}_{\ \ \ ;\,a;\,b}\ =\ -\Big({R}^c_{abm} +
{R}^c_{bma}\Big){^{\dagger}}\!{\Sigma}{^{abm}}\ =\ 
-{R}^c_{abm}{^{\dagger}}\! {\Sigma}{^{a[bm]}} -
{R}^c_{bam}{^{\dagger}}\! {\Sigma}{^{bam}}\ =
\\ \label{b26}
=\ -{R}^c_{abm}{^{\dagger}}\! {\Sigma}{^{a[bm]}} - 
{R}^c_{bam}{^{\dagger}}\! {\Sigma}{^{b[am]}}\ =\
-2{R}^c_{abm}{^{\dagger}}\! {\Sigma}{^{a[bm]}}.
\eey
Taking \R{b17a}$_2$ into account, we obtain
\bee{b27}
T^{bc}_{\ \ \ ;\,b}\ =\ -\frac{1}{2}{R}^c_{abm} S{^{abm}}.
\vspace{.3cm}\ee%%%%%%%%%%%%%%%%%%%%%%%%%%%%%%%%%%%%%
This demonstrates that the required compatibility of the relations \R{b5a}
with Einstein's field equations is guaranteed when the
Mathisson-Papapetrou equations \R{b17}$_3$ and \R{b27}are
satisfied\footnote[11]{These equations were first derived for pole-dipole
particles by Mathisson \C{9} and Papapetrou \C{10}. Later they were also proved
to be true for the free motion of continua with an intrinsic classical spin
$S^{abc}$, where $S^{abc}=u^a S^{bc}$. First time, this was done
by Weyssenhoff and Raabe \C{11} for the special model of an ideal fluid
with spin and, afterwards, for models specified by the choice of the
Lagrangian, as in \C{5,6}.}. In other words, for the sake of compatibility
the external force density and the external momentum density must be
spe\-ci\-fied as follows 
\bee{b28}
k^c\ :=\ -\frac{1}{2}{R}^c_{abm} S{^{abm}},\qquad
m^{[bc]}\ :=\ 2T^{[bc]}.
\vspace{.3cm}\ee%%%%%%%%%%%%%%%%%%%%%%%%%%%%%%%%%%%%%%%%%%%%%%
In continuum physics of SRT is common use, that energy-momentum and
spin can be balanced by \R{b5} and \R{b5o}. If one presupposes that in GRT the
general-covariantly written balances \R{b5a} of energy-momentum and spin
hold true ana\-lo\-gously to the SRT, then the external forces and the external
moments are specified by the gravitational field as given in \R{b28}:
energy-momentum and spin balances transfer into the Mathisson-Papapetrou
equations. The external sources do not vanish in GRT due to the gravitational
field, even if the energy-momentum tensor is symmetric. Consequently, the
Mathisson-Papapetrou equations are the basic equations of general-relativistic
continuum physics, if the validity of \R{b5a} is presupposed.

\end{document}